\documentclass[sigplan,nonacm]{acmart}

\usepackage{natbib}
\usepackage{url}
\usepackage{doi}
\usepackage{lipsum}

\usepackage{mathpartir}
\usepackage{hyperref}
\usepackage{graphicx}
\graphicspath{ {./images/} }

\usepackage{listings}
\newcommand{\bnfcomment}[1]{}

\newcommand{\fstE}[1]{\text{fst } #1}
\newcommand{\sndE}[1]{\text{snd } #1}
\newcommand{\sigmaE}[3]{\Sigma #1 : #2. \text{ } #3}
\newcommand{\unitE}{\langle \rangle}
\newcommand{\unittE}{\text{Unit}}
\newcommand{\univE}{\text{Type}}

\newcommand{\tyEqJ}[4]{#1 \vdash #2 = #3 : #4}
\newcommand{\stxEqJ}[3]{#1 \vdash #2 = #3}
\newcommand{\chkEqJ}[4]{#1 \vdash #2 = #3 \Leftarrow #4}
\newcommand{\synEqJ}[4]{#1 \vdash #2 = #3 \Rightarrow #4}
\newcommand{\freshJ}[2]{#1 \vdash #2 \text{ fresh}}
\newcommand{\steps}[2]{#1 \Downarrow #2}

\newcommand{\subst}[3]{#1 [#2 \mapsto #3]}

\newcommand{\mycomment}[1]{}

\usepackage[nameinlink]{cleveref}
\usepackage{framed}

\title{Extended Abstract: Towards a Performance Comparison of Syntax and Type-Directed NbE}
\author{Chester J.F. Gould}
\affiliation{%
  \institution{University of British Columbia}
  \city{Vancouver}
  \country{Canada}
}
\orcid{0009-0000-6365-7171}
\email{chester.gould@ubc.ca}

\author{William J. Bowman}
\affiliation{%
  \institution{University of British Columbia}
  \city{Vancouver}
  \country{Canada}
}
\orcid{0000-0002-6402-4840}
\email{wjb@williamjbowman.com}

\begin{document}

\begin{abstract}
        A key part of any dependent type-checker is the method for checking whether two types are equal.
A common claim is that syntax-directed equality is more performant, although type-directed equality is more expressive.
However, this claim is difficult to make precise, since implementations choose only one or the other approach, making a direct comparison impossible.
We present some work-in-progress developing a realistic platform for direct, apples-to-apples, comparison of the two approaches, quantifying how much slower type-directed equality checking is, and analyzing why and how it can be improved.

\end{abstract}

\maketitle

\section{Introduction}
Dependently typed languages treat types as first class values, enabling computation in types.
Consider the following.
{\small
\begin{verbatim}
int_or_str : (b : Bool) -> (if b then Int else String)
int_or_str = \b -> if b then 3141 else "Hello, World!"
\end{verbatim}
}
When applied to \texttt{true}, it will return the \texttt{Int} 3141.
When applied to \texttt{false}, it will return the \texttt{String "Hello, World!"}.
We can even express such facts as types.
{\small
\begin{verbatim}
int_or_str_theorem : int_or_str(true) = 3141
int_or_str_theorem = reflexivity
\end{verbatim}
}

Type-checking these requires deciding equality between two expressions.
The typing judgement $\Gamma \vdash e : A$ (expression $e$ has type $A$ with free variables $\Gamma$), must reason about equality of expressions.
In the true branch of \verb|int_or_str| we make the judgement $\vdash 3141 : \\({\small\verb|if true then Int else String|})$, and must reason that this type is equal to {\small\verb|Int|}.

Equality is captured by judgements of the form $\Gamma \vdash e_1 = e_2 : A$ ($e_1$ is equal to $e_2$ at type $A$ with free variables $\Gamma$).
The two judgements interact through the following typing rule.
\begin{mathpar}
  \inferrule*[left=Conv]{\Gamma \vdash e : B \\ \Gamma \vdash A = B : \texttt{Type}}{\Gamma \vdash e : A}
\end{mathpar}
An expression $e$ of type $B$ can also be considered of type $A$ so long as $A$ and $B$ are equal.

How to implement decision procedures for judgements of this form is an oft debated topic.

Our example relies on the following two rules.
\begin{mathpar}
  \inferrule*[left=App-$\beta$]
    {
    }
    {\Gamma \vdash (\text{\textbackslash} x \to b) a = \subst{b}{x}{a} : A
    }

  \inferrule*[left=if-true-$\beta$]
    {
    }
    {\Gamma \vdash \text{if true then } a \text{ else } b = a : A
    }
\end{mathpar}
Both are $\beta$ rules, which capture the execution of a program.
App-$\beta$ is the familiar $\lambda$-calculus $\beta$ rule, where a function literal applied to an argument is equal to the body of the function with the argument substituted for the parameter variable.
These rules are syntax-directed; we can see the type $A$ is not relevant in either rule. 

Another group of rules, called $\eta$ rules, give us additional equalities that rely on the type, rather than capturing the execution of a program.
\begin{mathpar}
  \inferrule*[left=Fun-$\eta$]
  { \Gamma, x : A \vdash f \hspace{0.2em} x = g \hspace{0.2em} x : B
  }
  { \Gamma \vdash f = g : (x : A) \to B
  }

  \inferrule*[left=Unit-$\eta$]
  {
  }
  { \Gamma \vdash a = b : \text{Unit}
  }
\end{mathpar}
The Fun-$\eta$ rule looks similar to the function extensionality principle and says that two functions are definitionally equal if they are definitionally equal when applied to an abstract variable.
The Unit-$\eta$ rule implies the Unit type has the single element $\unitE$.
Since there is only one element of the Unit type, all expressions of Unit type are equal.
This rule lets us type-check the following program.
{\small
\begin{verbatim}
unit_contractible : (x : Unit) -> (y : Unit) -> x = y
unit_contractible x y = reflexivity
\end{verbatim}
}

As \citet{Abel2013} shows, a widely used approach for deciding judgemental equality is normalization by evaluation (NbE).
An NbE algorithm maps the syntax of a language into a semantic domain and then back into syntax.
Since judgementally equal pieces of syntax will be equal in the semantic domain, going there, then back into syntax makes them syntactically equal (a process referred to as ``normalization").
NbE uses this process to turn the problem of deciding judgemental equality into one of giving semantics to syntax and then deciding syntactic equality.

Simple NbE algorithms implement an environment passing interpreter such as \citet{Coquand1996} or \citet{Chapman2005}.
These interpreters can have performance competitive with the widely used dependently typed languages, as \citet{smalltt} demonstrates with smalltt, a small but realistic, and very performant, type-checker for a dependent type theory.




However, another design decision remains: how to handle the type-directed $\eta$ rules (if at all).

Claims abound that syntax-directed approaches are more performant than those that integrate type-directed rules.
But how much more performant, in what situations, and is that performance worth the loss in expressivity?
It could be the case that handling type-directed rules causes untenable performance loss, or requires unwieldy implementation, meaning theories which require them should be avoided.
On the other hand, perhaps handling type-directed rules has only minimal performance cost and is quite simple to implement.
The problem is that no apples-to-apples comparison between type-directed and syntax-directed approaches has been made, and so there is currently no good way to decide between them beyond consulting a developer who has tried both, in an unknown context, and applying their subjective judgement to your context.
We would like to make these folklore claims more precise, and enable better informed choices.

\section{Benchmarking Platform}
Our approach is to compare two versions of smalltt~\cite{smalltt}.
The existing version uses a syntax-directed approach with $\eta$ rules for functions.

We implement two modified versions of smalltt with type-directed rules, trying to stay true to the performance consideration of smalltt.
The first merely transitions to a type-directed approach, while the second extends the type theory with $\eta$ for dependent pairs ($\Sigma$) and Unit.
Our two modified versions are available at \url{https://github.com/ChesterJFGould/smalltt} on the \texttt{master} and \texttt{sigma-unit} branches respectively.

To implement the type-directed algorithm, we follow the approach of \citet{Chapman2005}, who provide a systematic approach to derive an algorithm that supports type-directed rules by giving the type of the two normalized terms as an argument to the procedure deciding their equality.
The procedure can then inspect the type when it comes across a situation in which a type-directed rule is applicable.
This algorithm uses an approach similar to bidirectional typing~\cite{Dunfield2021} to reduce the amount of type information that is carried through the equality judgement.
\citet{Chapman2005} divide equality into two judgements: the check judgement $\chkEqJ{\Gamma}{n}{n'}{A}$, which takes a type as an input and in which $\eta$ rules can be implemented, and the synth judgement $\synEqJ{\Gamma}{\nu}{\nu'}{A}$ which checks two neutral forms for equality and additionally outputs their type.
We give an excerpt of the type-directed equality rules in \Cref{fig:unit-dependent-pair-type-directed-equality}.
In these judgements, the metavariable $n$ indicates terms that are normal with respect to $\beta$ rules, while $\nu$ indicates a neutral term, that is, a series of destructors applied to a free variable.

\begin{figure}[!htb]
  \begin{mathpar}
    \inferrule*[left=$\Sigma$-$\eta$]
    { \steps{\fstE{n}}{n_f} \\
      \steps{\fstE{n'}}{n_f'} \\
      \chkEqJ{\Gamma}{n_f}{n_f'}{n_1} \\
      \steps{\subst{e_2}{x}{n_f}}{n_2} \\
      \steps{\sndE{n}}{n_s} \\
      \steps{\sndE{n'}}{n_s'} \\
      \chkEqJ{\Gamma}{n_s}{n_s'}{n_2}
    }
    { \chkEqJ{\Gamma}{n}{n'}{\sigmaE{x}{n_1}{e_2}}
    }

    \inferrule*[left=$\Sigma$-T${=}$]
    { \chkEqJ{\Gamma}{n_1}{n_1'}{\univE}\qquad
      \freshJ{\Gamma}{y}\qquad
      \steps{\subst{e_2}{x}{y}}{n_2} \\
      \steps{\subst{e_2'}{x'}{y}}{n_2'} \\
      \chkEqJ{\Gamma}{n_2}{n_2'}{\univE}
    }
    { \chkEqJ{\Gamma}{\sigmaE{x}{n_1}{e_2}}{\sigmaE{x'}{n_1'}{e_2'}}{\univE}
    }

    \inferrule*[left=$\Sigma$-E1${=}$]
    { \synEqJ{\Gamma}{\nu}{\nu'}{\sigmaE{x}{n_1}{e_2}}
    }
    { \synEqJ{\Gamma}{\fstE{\nu}}{\fstE{\nu'}}{n_1}
    }

    \inferrule*[left=$\Sigma$-E2${=}$]
    { \synEqJ{\Gamma}{\nu}{\nu'}{\sigmaE{x}{n_1}{e_2}} \\
      \steps{\subst{e_2}{x}{\fstE{\nu}}}{n_2}
    }
    { \synEqJ{\Gamma}{\sndE{\nu}}{\sndE{\nu'}}{n_2}
    }

    \inferrule
    {
    }
    { \chkEqJ{\Gamma}{n}{n'}{\unittE}
    }
    \qquad
    \inferrule
    {
    }
    { \chkEqJ{\Gamma}{\unittE}{\unittE}{\univE}
    }
  \end{mathpar}
  \caption{Unit and Dependent Pair Type Directed Equality}
  \label{fig:unit-dependent-pair-type-directed-equality}
\end{figure}


In \Cref{fig:smalltt}, we present an overview of key definitions from smalltt that implement judgmental equality, and the changes we made to implement type-directed equality.
We have simplified the code shown here slightly to elide meta variables and unification, which are irrelevant to our purposes, but the implementation supports these. 

\begin{figure*}
\small
\begin{minipage}[t]{.47\textwidth}
\begin{minipage}[t]{.44\textwidth}
\begin{lstlisting}
data Tm = LocalVar Ix
  | App Tm Tm Icit
  | Lam NameIcit Tm
  | Pi NameIcit Ty Ty
  | U

type Ty = Tm

type Ix = Int

type Lvl = Int
\end{lstlisting}
\end{minipage}
\begin{minipage}[t]{.59\textwidth}
\begin{lstlisting}
data Val = VLocalVar Lvl Spine
  | VLam NameIcit Closure
  | VPi NameIcit VTy Closure
  | VU

type VTy = Val

data Spine = SId
  | SApp Spine Val

data Env = ENil
  | EDef Env Val
\end{lstlisting}
\end{minipage}
\begin{lstlisting}
data Closure = Closure Env Tm

eval :: Env -> Tm -> Val

unify :: Lvl -> Val -> Val -> IO ()

unifySp :: Lvl -> Spine -> Spine -> IO ()
\end{lstlisting}
\end{minipage}
\begin{minipage}[t]{.52\textwidth}
\begin{leftbar}
\begin{minipage}[t]{.46\textwidth}
\begin{lstlisting}
data Tm = LocalVar Ix
  | App Tm Tm Icit
  | Lam NameIcit Tm
  | Pi NameIcit Ty Ty
  | Sigma NameIcit Ty Ty
  | SigmaI Tm Tm
  | Fst Tm
  | Snd Tm
  | Unit
  | UnitI
  | U

type TypeCxt = Map Lvl VTy
\end{lstlisting}
\end{minipage}
\begin{minipage}[t]{.46\textwidth}
\begin{lstlisting}
data Val = VLocalVar Lvl Spine
  | VLam NameIcit Closure
  | VPi NameIcit VTy Closure
  | VSigma NameIcit VTy Closure
  | VSigmaI Val Val
  | VUnit
  | VUnitI
  | VU

data Spine = SId
  | SApp Spine Val
  | SFst Spine
  | SSnd Spine
\end{lstlisting}
\end{minipage}
\begin{lstlisting}
type Cxt = Cxt {lvl :: Lvl, localTypes :: TypeCxt}

unifyChk :: Cxt -> Val -> Val -> VTy -> IO ()

unifySp :: Cxt -> VTy -> Spine -> Spine -> IO VTy
\end{lstlisting}
\end{leftbar}
\end{minipage}
\caption{syntax-directed smalltt, key definitions (left); type-directed smalltt, key definitions (right)}
\label{fig:smalltt}
\end{figure*}

On the left, the \lstinline{Tm} type corresponds to our expressions, \lstinline{Val} to normal forms, and \lstinline{Spine} to the neutral terms.
The \lstinline{Val} type uses de Bruijn levels to represent variables instead of names.
For example, with de Bruijn levels, the term $\lambda x. \lambda y. x$ is represented as $\lambda. \lambda. 0$.
The \lstinline{Env} type represents a group of substitutions, and so the \lstinline{eval} function corresponds to a combination of the $\steps{e}{n}$ relation and the $\subst{e}{x}{e'}$ function.
A \lstinline{Closure} represents an \lstinline{Env} applied to a term.
Finally, \lstinline{unify} and \lstinline{unifySp} correspond to the syntax-directed equality judgment $\stxEqJ{\Gamma}{n}{n'}$, which elides types, but uses the current de Bruijn level to generate fresh variables instead of $\Gamma$.
Both \lstinline{unify} and \lstinline{unifySp} return an \lstinline{IO ()} since they will throw an exception if the terms aren't equal or evaluate to \lstinline{return ()} if they are.

On the right, we present the modifications made to convert smalltt to implement type-directed equality.
We only needed to change the type of \lstinline{unifyChk} and \lstinline{unifySp}.
\lstinline{unifyChk} now corresponds to the $\chkEqJ{\Gamma}{n}{n'}{A}$ judgment, while \lstinline{unifySp} now corresponds to the $\synEqJ{\Gamma}{\nu}{\nu'}{A}$ judgment.
We also add the unit and dependent pair types to our type-directed version of smalltt.

\section{Results and Conclusion}

\begin{figure}[!htb]
  \includegraphics[totalheight=16em]{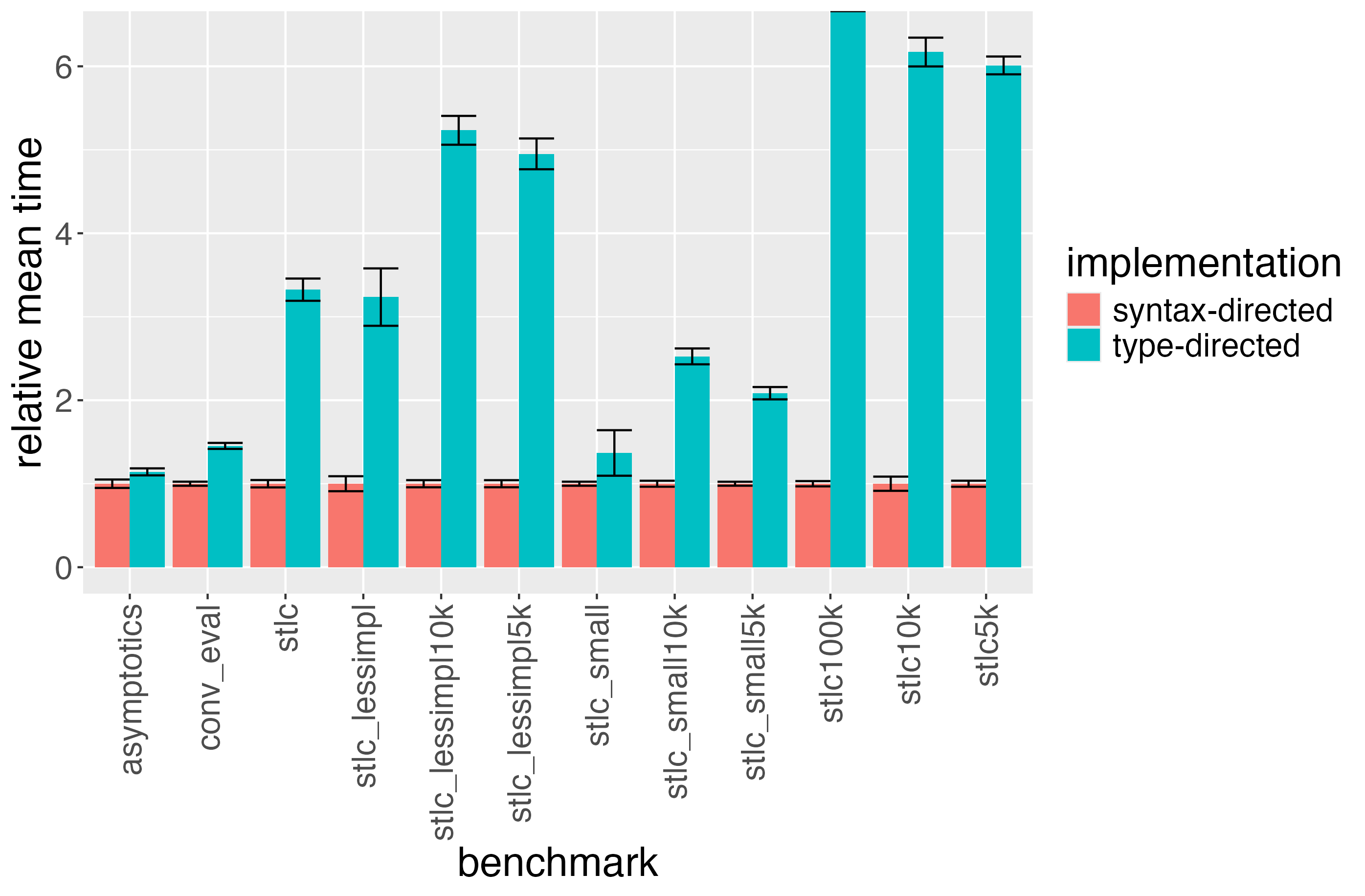}
  \caption{Benchmark Results}
  \label{fig:results}
\end{figure}

In \autoref{fig:results} we can see the results of the performance comparison between the original syntax-directed smalltt and our modified type-directed smalltt.
Each benchmark was run ten times on an AMD Ryzen 9 3900x processor with 16G of memory.
The results for each benchmark are presented so that the times for the type-directed implementation are given as a multiple of the syntax-directed times, with the syntax-directed times normalized to have a mean of 1.

Overall, the type-directed implementation performs worse, being on average 3.4 times slower than the syntax-directed implementation (excluding the benchmarks which didn't finish).
In the case of the stlc100k benchmark, the type-directed implementation failed to complete due to running out of memory.
We represent this as a column which continues off the top of the graph to infinity.
In comparison, benchmark numbers taken from equivalent benchmark suites show that Agda and Lean respectively perform 80 and 38 times slower on average than the syntax-directed smalltt implementation \citep{smalltt}.

We conjecture that the main reason for the discrepancy between the syntax and type-directed implementations, is that the syntax-directed implementation is able to take greater advantange of glued evaluation.
Glued evaluation exploits the fact that both substitution and evaluation respect judgmental equality to speed up the judgmental equality check.
In other words, if $\tyEqJ{\Gamma, x : A}{e}{e'}{B}$ then both $\tyEqJ{\Gamma}{\subst{e}{x}{d}}{\subst{e'}{x}{d}}{B}$ and $\tyEqJ{\Gamma}{n}{n'}{B}$ where $\steps{e}{n}$ and $\steps{e'}{n'}$.
Using these facts, we can avoid normalizing or performing a substitution on terms we are checking for equality if they are already judgmentally equal.

Glued evaluation comes to a head when comparing terms whose types contain many redexes.
Since we can only compare terms with the same type for equality, we must first compare the types of the terms, and glued evaluation lets us perform this comparison without computing many of the redexes.
In the type-directed implementation, however, we still fully normalize the type of both terms, since we need to know if it is a type at which we can apply an $\eta$ rule.

This conjecture leads us to the prediction that the performance difference between the syntax and type-directed implementations should correlate with the complexity of types used in the benchmark.
Indeed, we find that the benchmark for which the syntax and type-directed implementations are closest is the asymptotics benchmark which contains very little type-level computation, while the difference for the stlc benchmarks, which use quite complicated types to encode the syntax of the simply typed $\lambda$-calculus, is much larger.

As future work, we would like to further investigate this conjecture, and if it proves to be true, investigate if we can improve the performance of the type-directed approach.

Finally, there are other procedures for deriving a type-directed algorithm, and we'd like to integrate these into our comparison.
For example, in \citet{elabzoo}, the type of each term is calculated during normalization and then stored with the normalized term, so it can be inspected as needed during the equality check.
This approach may have different performance characteristics, but only supports the type-directed rule for Unit, using a syntax-directed method for $\Sigma$.
It's unclear whether either approach scales to other rules that require types, such as coproducts~\cite{altenkirch2001}.

%
%
%

\section{Acknowledgements}

\newcommand{\mynum}{RGPIN-2019-04207}







  We're thankful for the helpful feedback from the anonymous reviews.

  We acknowledge the support of the \grantsponsor{GS501100000023}{Natural Sciences and Engineering Research Council of Canada (NSERC)}{http://dx.doi.org/10.13039/501100000023}, funding reference number \grantnum{GS501100000023}{\mynum}.

  Cette recherche a été financée par le \grantsponsor{GS501100000023}{Conseil de recherches en sciences naturelles et en génie du Canada (CRSNG)}{http://dx.doi.org/10.13039/501100000023}, numéro de référence \grantnum{GS501100000023}{\mynum}.









\bibliography{main}

\end{document}